\documentclass[sigconf, screen]{acmart}

\ifpdf\usepackage{cmap}\fi

\usepackage{tcolorbox}
\usepackage{tabularx}

\tcbuselibrary{breakable} 

\AtBeginDocument{%
  }

\setcopyright{acmlicensed}
\copyrightyear{2025}
\acmYear{2025}
\acmDOI{XXXXXXX.XXXXXXX}
\acmConference[SIGSPATIAL '25]{33rd ACM SIGSPATIAL International Conference on Advances in Geographic Information Systems}{November 2025}{Minneapolis, MN, USA}
\acmISBN{978-1-4503-XXXX-X/2018/06}

\acmSubmissionID{123-A56-BU3}

\begin{document}

\title{TrajSceneLLM: A Multimodal Perspective on Semantic GPS Trajectory Analysis}

\author{Chunhou Ji}
\email{cji201@connect.hkust-gz.edu.cn}
\affiliation{%
    \institution{The Hong Kong University of Science and Technology (Guangzhou)}
    \city{Guangzhou}
    \country{China}
}

\author{Qiumeng Li}
\authornote{Corresponding author}
\email{qiumengli@hkust-gz.edu.cn}
\affiliation{%
    \institution{The Hong Kong University of Science and Technology (Guangzhou)}
    \city{GuangZhou}
    \country{China}
}

\renewcommand{\shortauthors}{Ji et al.}

\begin{abstract}
GPS trajectory data reveals valuable patterns of human mobility and urban dynamics, supporting a variety of spatial applications. However, traditional methods often struggle to extract deep semantic representations and incorporate contextual map information. We propose TrajSceneLLM, a multimodal perspective for enhancing semantic understanding of GPS trajectories. The framework integrates visualized map images (encoding spatial context) and textual descriptions generated through LLM reasoning (capturing temporal sequences and movement dynamics). Separate embeddings are generated for each modality and then concatenated to produce trajectory scene embeddings with rich semantic content which are further paired with a simple MLP classifier. We validate the proposed framework on Travel Mode Identification (TMI), a critical task for analyzing travel choices and understanding mobility behavior. Our experiments show that these embeddings achieve significant performance improvement, highlighting the advantage of our LLM-driven method in capturing deep spatio-temporal dependencies and reducing reliance on handcrafted features. This semantic enhancement promises significant potential for diverse downstream applications and future research in geospatial artificial intelligence. 
The source code and dataset are publicly available at:\href{https://github.com/februarysea/TrajSceneLLM}{https://github.com/februarysea/TrajSceneLLM}.
\end{abstract}

\ccsdesc[500]{Computing methodologies~Artificial intelligence}
\ccsdesc[500]{Human-centered computing~Geographic visualization}

\keywords{Travel Mode Identification, GPS Trajectory, Geospatial Data, Large Language Models, Geospatial Context}


\begin{teaserfigure}
  \includegraphics[width=\textwidth]{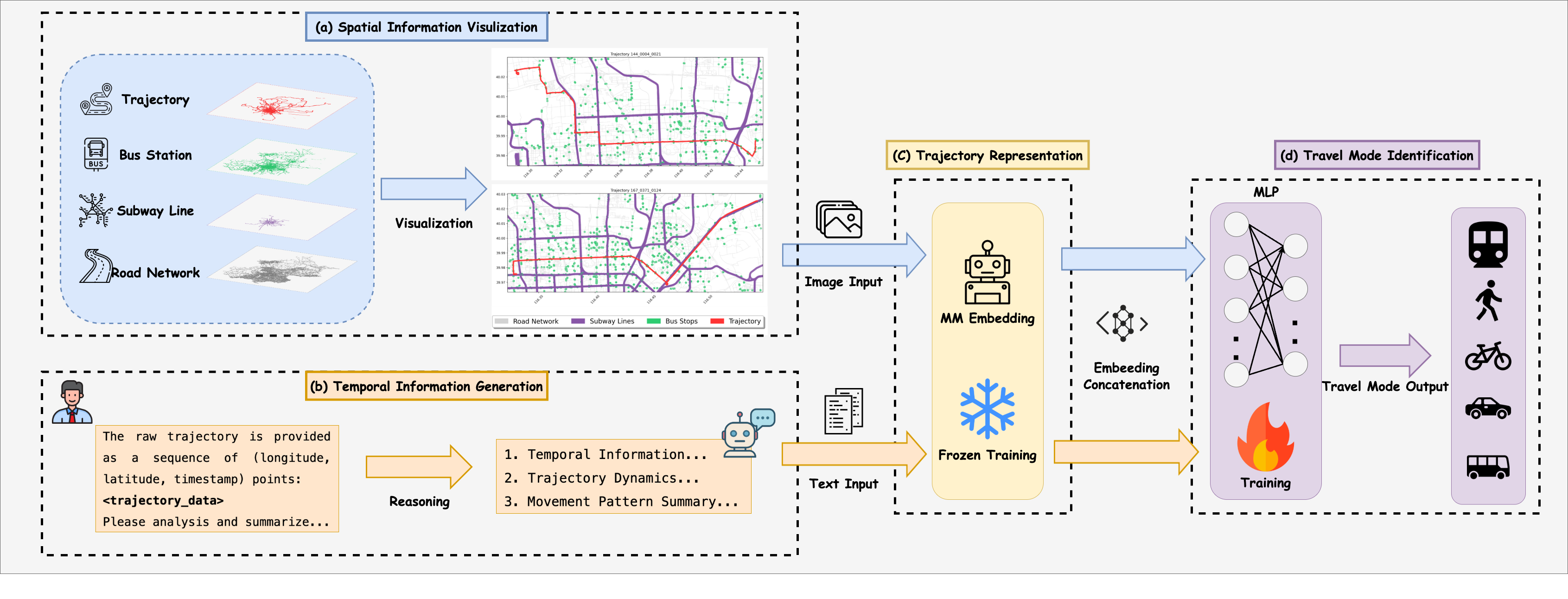}
  \caption{A multimodal LLM-based framework for trajectory scene understanding, comprising: (a) spatial information generation through trajectory-map visualization; (b) temporal feature extraction by an LLM; (c) embedding generation by fusing visual and textual inputs via a multimodal LLM; and (d) travel mode identification using an MLP classifier.}
  \Description{A four-stage framework diagram. Stage A shows map and trajectory input. Stage B shows textual data generation. Stage C shows multi-modal embedding. Stage D shows travel mode output. Arrows indicate data flow from GPS trajectory input to travel mode output.}
  \label{fig:teaser}
\end{teaserfigure}


\maketitle

\section{Introduction}
Trajectory data provides valuable insights into human mobility and urban dynamics, with GPS feeds offering detailed spatio-temporal information important for such analysis\cite{gao2015spatio, chen2024deep}
However, existing analysis methods often struggle to mine deep semantics from raw trajectory points, especially in effectively integrating and understanding their complex multidimensional contexts\cite{zhou2024red,fu2020trembr}.
The conventional data processing in such methods, often focused on creating structured representations, restricts their ability to incorporate intuitive visual information from map scenes, and they also frequently lack effective unified representations for diverse contextual data.

Recent studies have sought to address these limitations from various angles. Some of them focus on enriching trajectories by integrating diverse semantic categories \cite{qian2024context, dabiri2018inferring,yin2021gps2vec}.
Others explore leveraging image-based representations for trajectory processing \cite{yutian2025mvcf,ma2023multi,ribeiro2024deep}.
However, these approaches often need sophisticated feature engineering, including the extraction of specific trajectory characteristics.
Meanwhile, the adoption of large language models (LLMs) for trajectory mining is gaining traction\cite{liu2024semantic, ji2024evaluating, zhang2024large}, though current applications remain limited in their ability to model complex multimodal contexts\cite{lu2023theory}. Multimodal LLMs are well suited to this challenge. Designed to process and fuse diverse input types, they have shown strong semantic reasoning capabilities in other domains\cite{achiam2023gpt, team2023gemini, guo2025seed1, guo2025deepseek}. However, their potential remains underexplored in GPS trajectory analysis—particularly in achieving a unified, semantic interpretation by integrating visualized map scenes with temporally rich and dynamic textual descriptions.
We define the trajectory scene as a multimodal representation that combines a GPS trajectory’s visual-spatial configuration and its semantic-dynamic context. 

In this work, we propose a new perspective on GPS trajectory analysis, one that utilizes multimodal LLMs to comprehensively enhance the semantics of GPS trajectories within their spatio-temporal environments.
This is achieved by first generating distinct semantic embeddings from two complementary modalities, and then concatenating them. The two modalities are: 1) visualized map images that intuitively present the trajectory's spatial configuration alongside key geographical context, and 2) textual contexts that convey LLM-summarized trajectory dynamics and precise temporal information.
These embeddings are then concatenated to produce a final trajectory scene representation, which proves to be rich and highly discriminative for the downstream task.
The efficacy of these embeddings and our overall perspective is validated through the downstream task of TMI using the GeoLife dataset\cite{zheng2010geolife, zheng2008understanding, zheng2009mining}. We selected TMI due to its foundational role in analyzing urban transport systems. Accurately identifying the mode of transport is the first step towards understanding individual travel choices and modeling large-scale mobility behaviors,
The main contributions of our work include:
\begin{itemize}
    \item Proposing a new perspective for GPS trajectory analysis using multimodal LLMs for semantic understanding.
    \item Introducing a new method that leverages  reasoning LLM to extract and summarize key spatio-temporal characteristics from GPS trajectories, forming a rich textual input for multi-modal analysis.
    \item Providing initial evidence for the potential of the proposed perspective through its application to enhance TMI.
\end{itemize}

\section{Methodology}

The framework based on our perspective(\hyperref[fig:teaser]{Figure~\ref*{fig:teaser}}) processes raw GPS trajectories to generate enhanced semantic representations of the 'trajectory scene'.

\subsection{Multi-modal Input Generation}

A raw GPS trajectory point $p_i$ is defined by its geographical coordinates (longitude $lon_i$ and latitude $lat_i$) and its timestamp $ts_i$:
\begin{equation}
    p_i=(lon_i,lat_i,ts_i)
\end{equation}
A trajectory segment, $Tr$, is a chronologically ordered sequence of $k$ such points:
\begin{equation}
    Tr=\langle p_1,p_2,...,p_k \rangle
\end{equation}
These segments $Tr$, serving as common input for generating both visual and textual modalities, undergo necessary data cleaning procedures before processing.

\subsubsection{Spatial Information Visualization}
This stage (\hyperref[fig:teaser]{Figure~\ref*{fig:teaser}(a)}) generates visualized map images from each trajectory segment $Tr$. Each image displays $Tr$ integrated with key contextual map layers including the road network, subway lines, and bus stations, with its geographical extent dynamically determined by $Tr$'s spatial span plus an added buffer to include relevant surrounding context. 
An example visualization, which also illustrates the color scheme used for clarity (e.g., $Tr$ in red), is provided in \hyperref[app:example_inputs]{Appendix~A}( \hyperref[fig:figure2]{Figure~\ref*{fig:figure2}}). Geospatial data for the map layers are obtained from OpenStreetMap (OSM)\cite{OpenStreetMap}.

\subsubsection{Temporal Information Generation}

This stage (\hyperref[fig:teaser]{Figure~\ref*{fig:teaser}(b)}) employs reasoning-LLM DeepSeek-R1\cite{guo2025deepseek}, proficient in complex reasoning, to generate rich textual information. 
For each trajectory segment $Tr$, the reasoning-LLM uses a designed prompt to analyze raw data and extract key Temporal Information (e.g., start/end times, duration, inactivity periods) and Trajectory Dynamics (e.g., speed profiles, turn frequency, detour index). 
The LLM provides both these detailed extracted features and a synthesized summary of the overall movement pattern and temporal characteristics.
A comprehensive example of all extracted textual features and the summary is available in \hyperref[app:example_inputs]{Appendix~A}.

\subsection{Trajectory Scene Representation}

The visualized map images and textual contexts serve as complementary inputs for generating representations.  We employ the multimodal LLM, Seed1.5-VL\cite{guo2025seed1}, to generate separate embeddings for each of these modalities(\hyperref[fig:teaser]{Figure~\ref*{fig:teaser}(c)}). These two embeddings are then concatenated to produce the final representation, the semantic embedding of the trajectory scene.

For the downstream task of TMI (\hyperref[fig:teaser]{Figure~\ref*{fig:teaser}(d)}), these semantic embeddings are then fed as input features to a simple MLP classifier. 
This MLP focuses on efficiently distinguishing travel modes from these rich embeddings, thereby balancing representation strength with downstream efficiency.
While the multimodal LLM could perform end-to-end classification, we decouple its role to focus on comprehensive multi-modal representation learning, thereby generating versatile trajectory scene embeddings.
These embeddings can effectively inform various separate, potentially simpler, classification models (like the MLP for TMI) and are designed to be adaptable for other trajectory-related analytical tasks, thus promoting broader applicability.

\section{Experiment Results}
To validate the effectiveness of our proposed TrajSceneLLM framework, we conducted extensive experiments for the task of TMI on the GeoLife dataset\cite{zheng2010geolife, zheng2008understanding, zheng2009mining}. Our analysis focused on five primary travel modes that are highly representative of urban mobility: walk, bike, bus, car, and subway. The experimental evaluation involved comparisons against several baseline methods, including traditional models (SVM, RF) trained on handcrafted features, the state-of-the-art semantic feature learning model MASO-MSF\cite{ma2023multi}, and a 'direct' baseline using raw LLM embeddings to assess their intrinsic representational capacity. Furthermore, we performed ablation studies to analyze two key aspects: the individual contribution of each modality (by removing either the visual or textual input) and the effectiveness of different embedding combination strategies (concatenation vs. fusion).

Our proposed TrajSceneLLM framework, using concatenated embeddings, achieves a state-of-the-art accuracy of 86.8\% on the TMI task. This result represents a significant improvement not only over traditional machine learning models like SVM (76.1\%) and RF (81.6\%), which are limited by their reliance on handcrafted features , but also over the strong semantic feature learning baseline, MASO-MSF (84.4\%). We attribute this 2.4\% performance gain over MASO-MSF to TrajSceneLLM's unique ability to integrate intuitive, visual map context with deep, LLM-reasoned temporal dynamics—a capability that existing methods lack.
To dissect the sources of this performance gain, we conducted a series of ablation studies, the results of which are detailed in Table 1. Critically, removing the visual modality (Ours w/o. image) led to a accuracy drop to 82.8\%. This underscores the importance of the spatial scene context; for instance, visual confirmation of a trajectory's alignment with a subway line provides a powerful and unambiguous feature that is difficult to capture through textual or kinetic data alone. Even more telling is the decline to 81.9\% accuracy upon removing the LLM-generated text (Ours w/o. text). This demonstrates the indispensable value of the nuanced temporal and dynamic summaries, which capture subtle differences in movement patterns that are essential for distinguishing between road-based transport modes. These findings collectively validate our central thesis: the synergistic combination of visual-spatial and textual-temporal modalities provides a richer, more discriminative trajectory representation, moving beyond raw data to genuine semantic understanding.

\begin{table}[htbp]
  \centering
  \small
  \caption{Comparison of Different Model Configurations for Travel Mode Identification.}
  \label{tab:model_comparison_full_metrics}
  \begin{tabularx}{\columnwidth}{@{} l >{\raggedright\arraybackslash}X cccc @{}} 
    \toprule
Model & Acc (\%) & Precision (\%) & Recall (\%) & F-Score (\%) \\ 
    \midrule
    SVM & 76.1 & 76.0 & 65.9 & 68.7 \\
    RF & 81.6 & 80.0 & 74.0 & 77.6 \\
    MASO-MSF & 84.4 & 82.3 & 81.2 & 81.3 \\
    \midrule
    Direct(w/o. text) & 38.8 & 44.8 & 38.8 & 34.9\\ 
    Direct(w/o. image) & 48.2  & 58.8 & 48.2 & 46.6 \\ %
    Direct & 48.5 & 58.3 & 48.5 & 47.8 \\   
    \midrule
    Ours(w/o. text) & 81.9 & 84.5 & 75.8 & 79.2 \\
    Ours(w/o. image)& 82.8 & 82.0 & 79.4 & 80.6 \\ 
    Ours(fusion) & 83.5 & 83.2 & 80.0 & 81.6\\ 
   \textbf{Ours(concatenation)} & \textbf{86.8} & \textbf{85.2} & \textbf{84.8} &  \textbf{85.0}\\ 
    \bottomrule
  \end{tabularx}
  \Description{A table comparing different models (grouped by type: Baseline, Direct, Proposed) on Accuracy, Precision, Recall, and F-Score for travel mode identification, filling the column width.} 
\end{table}

\section{Conclusion}
We introduce a new multimodal perspective on GPS trajectory understanding through semantic enrichment using large language models. 
Our proposed framework, TrajSceneLLM, integrates visualized map images with LLM-generated textual descriptions to produce semantically rich trajectory scene embeddings.
Experiments on the GeoLife dataset demonstrate that these embeddings enhance TMI performance when used with a MLP classifier.
Future work could build upon these findings like incorporating Point-of-Interest (POI) data to infer the semantic purpose of trips. Our results highlight the promising potential of multimodal LLMs in advancing semantic trajectory analysis and enabling a wide range of geospatial AI applications.

\bibliographystyle{ACM-Reference-Format}
\bibliography{myreferences}

\appendix
\label{app:example_inputs} 

\section{Example Inputs for the multimodal LLM}
The following example presents both the spatial information as a map and the corresponding LLM-generated temporal information for a sample trajectory segment.
 
\begin{figure}[htbp] 
    \centering 
    \includegraphics[width=0.8\linewidth]{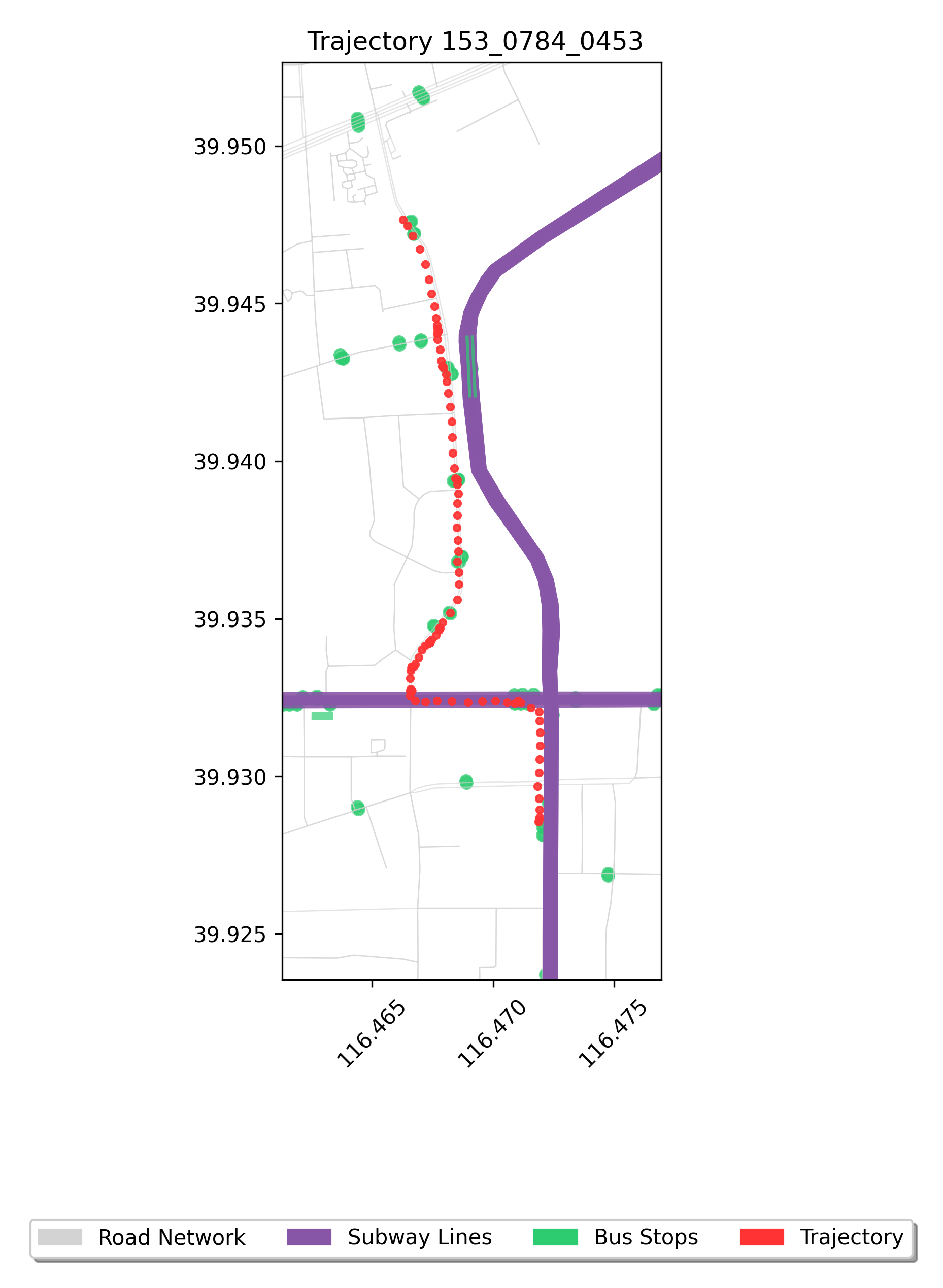} 
    \caption{Example Inputs for the Multimodal LLM}
    \label{fig:figure2} 
\end{figure}

\begin{tcolorbox}[
    breakable,
    colback=green!5!white,
    colframe=green!60!black,
    title=Temporal and dynamic information for the bus trajectory
]

\textbf{1. Temporal Information}\par
\smallskip
\textbf{Start/End Time}: The trajectory spans from \textbf{1236500298.0} (April 8, 2009, $\sim$15:38 UTC; local time $\sim$23:38 Beijing) to \textbf{1236500943.0} (April 8, 2009, $\sim$15:49 UTC; local time $\sim$23:49 Beijing).\par
\textbf{Total Duration}: $\sim$10.75 minutes (645 seconds).\par
\textbf{Day Type}: Occurred on a weekday (Wednesday, April 8, 2009).\par
\textbf{Time of Day}: Late evening/night (local time 23:38–23:49), outside peak commuting hours (typically 7–9 AM and 5–7 PM local time).\par
\textbf{Inactivity Periods}: Brief stops (2–8 seconds) detected at multiple points (e.g., 1236500363.0–1236500366.0, 1236500673.0–1236500681.0), likely due to GPS noise or momentary pauses.\par

\par\noindent\rule{\linewidth}{0.4pt}\par\nopagebreak 
\medskip 

\textbf{2. Trajectory Dynamics}\par
\smallskip
\textbf{Average Speed}: Estimated at $\sim$11 km/h (3.1 m/s), calculated as total travel distance ($\sim$2,000 meters) divided by total duration (645 seconds).\par
\textbf{Speed Variation}: Moderate fluctuations (5–29 km/h in initial segments), with deceleration during brief stops.\par
\textbf{Turn Frequency}: Minimal sharp turns; movement primarily follows a gradual, southeast-to-northwest path with small angular changes ($<$30° between consecutive segments).\par
\textbf{Stops}: Multiple brief stationary periods (2–8 seconds) but no prolonged stops ($>$10 seconds).\par
\textbf{Total vs. Straight-Line Distance}: Total travel distance $\approx$2,000 meters; straight-line distance $\approx$687 meters.\par
\textbf{Detour Index}: $\sim$2.9 (actual path length/straight-line distance), indicating a moderately winding route or detour.\par

\par\noindent\rule{\linewidth}{0.4pt}\par\nopagebreak
\medskip 

\textbf{Overall Movement Pattern Summary}
\medskip

This trajectory reflects a short ($\sim$10.75-minute) weekday trip during late evening, outside peak commuting hours. The movement is characterized by a moderate average speed ($\sim$11 km/h), consistent with cycling or slow motorized travel, with brief stops likely due to momentary pauses or GPS noise. The path is moderately winding (detour index $\sim$2.9), suggesting navigation through urban streets or a non-direct route. Speed variations and minimal sharp turns align with typical urban mobility patterns for non-motorized or slow-motorized modes.
\end{tcolorbox}

\end{document}